\documentclass{article}

\usepackage[final]{neurips_2020}

\usepackage[utf8]{inputenc} %
\usepackage[T1]{fontenc}    %
\usepackage{hyperref}       %
\usepackage{url}            %
\usepackage{booktabs}       %
\usepackage{amsfonts}       %
\usepackage{nicefrac}       %
\usepackage{microtype}      %
\usepackage{xcolor}         %
\usepackage[capitalize,noabbrev]{cleveref} %

\usepackage{graphicx}

\title{Like a Researcher Stating Broader Impact 
\newline For the Very First Time}

\author{%
  Grace Abuhamad \\
  Element AI \\
  Montreal, Canada \\
  \texttt{grace@elementai.com} \\
   \And
  Claudel Rheault \\
  Element AI \\
  Montreal, Canada \\
   \texttt{claudel.rheault@elementai.com} \\
}

\begin{document}

\maketitle

\begin{abstract}
  In requiring that a statement of broader impact accompany all submissions for this year’s conference, the NeurIPS program chairs made ethics part of the stake in groundbreaking AI research. While there is precedent from other fields and increasing awareness within the NeurIPS community, this paper seeks to answer the question of how individual researchers reacted to the new requirement, including not just their views, but also their experience in drafting and their reflections after paper acceptances. We present survey results and considerations to inform the next iteration of the broader impact requirement should it remain a requirement for future NeurIPS conferences.  
\end{abstract}

\section{Introduction}

There are increasing examples of unethical uses of technology [15, 21]. To curb this trend, some proposals attempt to limit investment or procurement without impact assessment [13] or call for outright bans [12, 24]. Others aim to instill ethical practices at the research stage, before technology transfers into products [14, 15]. Conferences that are “traditionally” technical have started to host workshops on social impact issues [2, 3, 4, 8, 16], and in some cases, announced more interdisciplinary subject areas [7]. 

Perhaps the most significant change is the NeurIPS submission requirement for a statement of broader impact [23]. Unlike workshops and interdisciplinary tracks that might have been considered more “niche”, the requirement directly affects every submission, of which there are over 9000 this year [19]. While broader impact statements in themselves are not new to the research community at-large [11, 25, 26], they are new to the NeurIPS community. This paper seeks to answer the question of how individual researchers reacted to the new requirement, including not just their views, but also their experience and process in drafting and their reflections on the requirement after paper acceptances. 

This research started as an internal discussion at our organization, and subsequently folded into a public conversation as interest grew on the topic. To collect researcher perspectives within and beyond our organization, we created an online public survey [5]. The results inform considerations for designing the next iteration of the broader impact requirement, should it remain in place for future NeurIPS conferences. While we recognize that researchers are not the only intended audience for broader impact statements and that they might not be the only ones with roles and responsibilities in ethical research and technology development, researchers represent a critical mass to mobilize in this effort. Understanding researchers’ experience and process is essential not just to the design of the NeurIPS requirement, but also to advancing towards the goal of ethical research practices in general, given that NeurIPS is a top-tier conference with influence effects beyond its annual meeting.

\section{Survey Method}

The method used was an exploratory mixed-methods survey with open and closed-ended questions, split in two main sections: section 1 for researchers who had submitted to NeurIPS and section 2 for those who had not. The survey was anonymous and no demographic information was collected.\footnote{The survey did not undergo assessment by an Institutional Review Board (IRB).} The survey was distributed online within research community channels (e.g., Slack, email lists) and via social media. The goals were to: 1) understand how researchers investigated the implications of their research and subsequently, how they defined their impact statement, as well as 2) understand their views on this new submission requirement. Survey questions focused first on the approach taken for writing the statement and on challenges encountered, then on the perceptions of influence their statement had on the overall submission, and finally, on how they felt about this newly added requirement. 

\section{Survey Results}

A total of 50 participants responded to the survey, most of whom identified themselves as academics (72\%), and industry researchers (23.5\%). There was a balanced breakdown by career stage, with the largest group of respondents identifying as graduate students (33\%). For the group who submitted to NeurIPS (74\% of respondents), the majority identified their subject areas as deep learning and theory, respectively. However, among researchers who did not submit to NeurIPS (26\%), the primary subject areas were deep learning and social aspects of machine learning. We did not compare our survey population to the NeurIPS population, though that may be an interesting area for future study. Our questions focused first on the process and challenges in completing the submission requirement, then on the perceived impact of the requirement on paper  acceptances, and finally on researchers’ views on the requirement and its framing. 

\subsection{Process and challenges}

\begin{figure}%
  \centering
  \includegraphics[width=\linewidth]{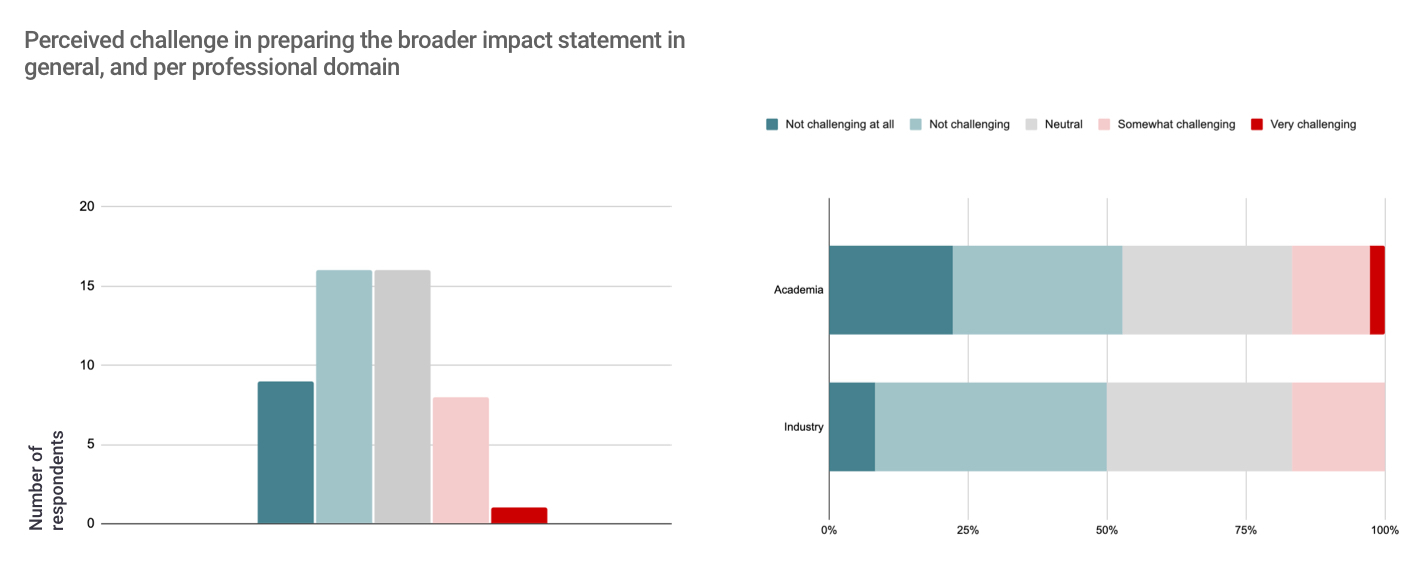}
  \caption{Level of perceived challenge in preparing the broader impact statement}
  \label{Figure 1}
\end{figure}

When asked about how they approached the completion of their broader impact statements, 83.8\% of respondents indicated that they completed this part within the co-authors group, without external help. The rest of participants adopted different approaches to completing it, for example accepting support or reaching out for help. A vast majority spent less than 2 hours on this part of their submission, but almost half of them mentioned it was not challenging or not challenging at all to prepare it. When asked about what could make this difficult, there appeared  to be different trends. Some viewed their theoretical work as so far from potential practical applications that they found the exercise speculative. Others perceived the requirement as a “bureaucratic constraint” and did not take it seriously. Perhaps not surprisingly, researchers at different stages of their career found the exercise more or less challenging, although their professional domain (academia or industry) does not appear to be a factor in their experienced difficulty with the exercise (\cref{Figure 1}). 

\subsection{Impact on submission}

\begin{figure}%
  \centering
  \includegraphics[width=\linewidth]{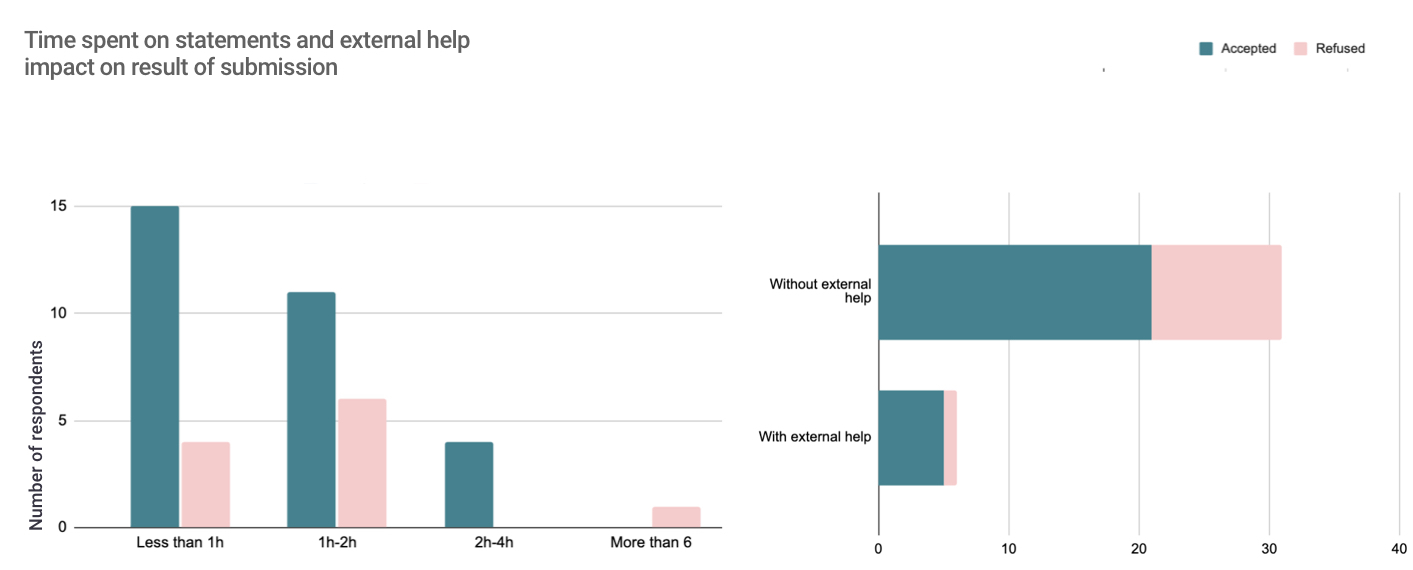}
  \caption{Impact of time spent and help received in preparing statements}
    \label{Figure 2}
\end{figure}

Although it was made clear that submissions would not be rejected solely on the basis of the broader impact statement [6], the survey explored researchers’ perspectives on this point. For researchers who submitted, more than 75\% said they believed the statements were not taken into consideration at all, yet almost 90\% thought that it was unclear exactly how reviewers would assess if the statements were adequately addressing the impacts. Surprisingly, even if the evaluation process was unclear, when asked how confident they were that their statement was adequately responding to the requirement, 43.2\% said they were either confident or very confident. Based on our results, time spent did not seem to have an impact, since most of the respondents who spent less than an hour on preparing their broader impact statement also received acceptances of their submissions. Those who sought external help appear to have a lower ratio of rejections (see \cref{Figure 2}), but we recognize that our sample size may be too small to draw conclusive results.

\subsection{Framing}

\begin{figure}%
  \centering
  \includegraphics[width=\linewidth]{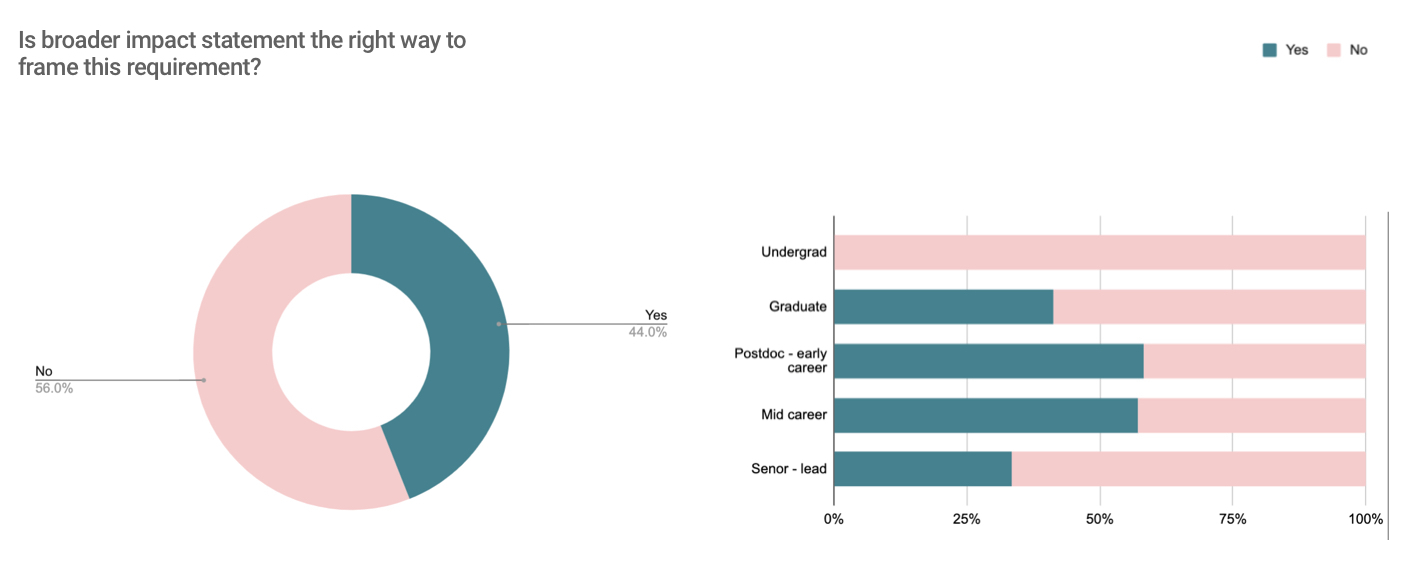}
  \caption{Respondents’ views on framing the requirement as a statement of broader impact}
  \label{Figure 3}
\end{figure}

The survey explored researchers’ views on the requirement and its framing. Our results indicated that the community is divided on how to frame the requirement: 56\% did not agree that broader impact was the right way to frame the requirement, while 44\% did (\cref{Figure 3}). This split was similar when compared to subject area (i.e., those who declared “theory” as one of their subject areas were equally split), submitters/non-submitters, and academia/industry. Postdoctoral/early-career and mid-career respondents were more supportive of the requirement framing than students and senior researchers. There seems to be a general feeling that assessing broader impact is important, but some uncertainty regarding who should do it and how. Some respondents described the requirement as “too broad” or said they did not feel “qualified to address the broader impact of [their] work.” Among those who supported the requirement, some found the thought process most valuable, and that it “forces researchers to reflect on the impact of their research.”

\section{Integrating feedback into next iteration of broader impact}

The survey results serve not only to examine researchers’ perspectives and formulation process but also to inform the next iteration of the broader impact requirement should it remain in place for future NeurIPS conferences. When asked what could have helped them most in the process of preparing their statement of broader impact, 92\% of respondents indicated that examples of broader impact statements would be most helpful. While a few examples were provided this year [6], there will of course be an increasing number of examples to draw from in future years. Guidelines were the second most popular request, including both guidelines on when a statement might be applicable or not as well as how to formulate a statement. NeurIPS did not provide specific guidelines (apart from referencing [16]), however a group of researchers published a guide [10] that may serve for future guideline development. This section proposes, based on the survey results and observations, how to integrate respondent feedback into future iterations of broader impact: rethinking the requirement design and framing; developing greater capacity and confidence among researchers; and reflecting the shared responsibility of ethical research and technology development in society. 

\subsection{Requirement design}

If the goal of the broader impact requirement is to develop ethical research practices, there may be other ways by which to achieve this goal. We recognize that there is precedent for written statements in grant writing [22] and that given the paper-based nature of submissions, a written statement may seem like the most logical implementation of such a requirement. However, our respondents indicated a combination of nonchalance in their approach to the requirement (“This is just another exercise in ‘doing something’. I expect these statements to become pro forma with time, since it will be possible to look at previous years' papers for ‘inspiration’.”; “I wasn't particularly rigorous about it.”), outright farce (“If I liked writing fiction I would be writing novels.”), or perceived it as a burden (“one more burden that falls on the shoulders of already overworked researchers”). It is possible that these attitudes may have a counterproductive effect on an ethical research goal [14]. We invite program chairs to consider mechanisms to limit that effect (e.g., an incentive for “best” broader impact statements [4]). Such mechanisms are important not only to manage negative effects, but also to encourage researchers who found the exercise valuable.  

\subsection{Capacity-building}

Given that many survey respondents felt they were not qualified to address the broader impact of their work, workshops [20, 25] may help build capacity over time, especially where workshops might provide a space for researchers to examine their work with a more diverse group of researchers (in terms of experience, demographics, discipline, etc.). Discussions of this type not only help develop capacity and confidence, but also may surface areas of impact that were overlooked by authors. Interdisciplinary collaborations could also introduce new guidelines or methodologies such as the theory of change [28] or consequence scanning [27].

\subsection{Shared responsibility}

Recognizing how different systems and social worlds are interacting would increase the quality of the discussion on broader impact, and, perhaps more importantly, develop a sense of shared responsibility within the ecosystem for ethical research and technology development [17]. Researchers represent a critical mass, but others, such as conference organizers, institutions, funders, and users also have roles and responsibilities [1, 9, 18, 25]. Perhaps as a way to address concerns around burden and requisite expertise, the assessment of broader impact could be more of a multistakeholder exercise. 

\section{Conclusion}

This paper and its underlying survey investigated how researchers approached the broader impact statement and surface considerations to better design the requirement in future years. While the survey represented a small sample of the NeurIPS community, its results demonstrate nonetheless a split regarding how the requirement is framed, and how important it seems to be for researchers. Initiating a conversation about broader impact is in itself already a step towards establishing norms and best practices for ethical research. We encourage further work to monitor the evolution of researchers’ perspectives not only at top-tier conferences such as NeurIPS, but also at-large.

\begin{ack}
The authors would like to thank Noémie Lachance, Tara Tressel, and the Element AI Research Group for their support and participation throughout the process (which started back in March 2020).
\end{ack}

\section*{Supplementary Material}
The survey questions \textcolor{blue}{\href{https://drive.google.com/file/d/15acTwJjfhQZrwaTrerXTR1pTkyvwrErD/view?usp=sharing}{(here)}} and the responses received \textcolor{blue}{\href{https://docs.google.com/spreadsheets/d/1PjLuregC1aeKO0JsPVNr-vXuCFcL4TrS6qcuaN8Mph4/edit?usp=sharing}{(here)}} are available for further investigation and use. The survey remains open to responses. At the time of drafting, we had 50 responses on which we conducted the analysis for this paper.

\section*{References}

\small

[1] “AI Algorithms Need FDA-Style Drug Trials” (2019) Wired. https://www.wired.com/story/ai-algorithms-need-drug-trials/. 
[2] “AI For Social Good” (2019). Neural Information Processing Systems Conference. https://aiforsocialgood.github.io/neurips2019/. 

[3] “Minding the Gap: Between Fairness and Ethics” (2019). Neural Information Processing Systems Conference. https://mindingthegap.github.io/. 

[4] “Navigating the Broader Impacts of AI Research” (2020). Neural Information Processing Systems Conference.  https://nbiair.com/index.html. 

[5] “NeurIPS 2020 - Feedback on the Broader Impact Statement” (2020) https://forms.gle/W9tdrBfM9oybh68YA. 

[6] “NeurIPS 2020 FAQ For Authors” (2020).  https://neurips.cc/Conferences/2020/PaperInformation/NeurIPS-FAQ.

[7]  “NeurIPS 2020 Subject Areas” (2020). https://nips.cc/Conferences/2020/PaperInformation/SubjectAreas. 

[8] “Workshop on Ethical, Social and Governance Issues in AI” (2018). Neural Information Processing Systems Conference. https://sites.google.com/view/aiethicsworkshop/the-workshop?authuser=0. 

[9] Askell, A., Brundage, M., \& Hadfield, G. (2019). The role of cooperation in responsible AI development. arXiv preprint arXiv:1907.04534.

[10] Ashurst, C.; Anderljung, M.; Prunkl, C.; Leike, J.; Gal, Y.; Shevlane, T.; Dafoe, A.. (2020). A Guide to Writing the NeurIPS Impact Statement. Centre for the Governance of AI. https://medium.com/@GovAI/a-guide-to-writing-the-neurips-impact-statement-4293b723f832

[11] Bornmann, L. (2013). What is societal impact of research and how can it be assessed? A literature survey. Journal of the American Society for information science and technology, 64(2), 217-233.

[12] Coalition for Critical Technology. 2020. “Abolish the \#TechToPrisonPipeline” Medium. https://medium.com/@CoalitionForCriticalTechnology/abolish-the-techtoprisonpipeline-9b5b14366b16. 

[13] Dawson, P. (2019). “Closing the Human Rights Gap in AI Governance” Element AI. https://www.elementai.com/news/2019/supporting-rights-respecting-ai. 

[14] Gupta, A., Lanteigne, C., \& Heath, V. (2020). Report prepared by the Montreal AI Ethics Institute (MAIEI) for Publication Norms for Responsible AI by Partnership on AI. arXiv preprint arXiv:2009.07262.

[15] Hecht, B., Wilcox, L., Bigham, J.P., Schöning, J., Hoque, E., Ernst, J., Bisk, Y., De Russis, L., Yarosh, L., Anjum, B., Contractor, D. and Wu, C. (2018). It’s Time to Do Something: Mitigating the Negative Impacts of Computing Through a Change to the Peer Review Process. ACM Future of Computing Blog. https://acm-fca.org/2018/03/29/negativeimpacts/.

[16] Hecht, B., (2020). “Suggestions for Writing NeurIPS 2020 Broader Impacts Statements” Medium. https://medium.com/@BrentH/suggestions-for-writing-neurips-2020-broader-impacts-statements-121da1b765bf. 

[17] Hervieux, C. and Voltan, A. (2019), "Toward a systems approach to social impact assessment", Social Enterprise Journal, Vol. 15 No. 2, pp. 264-286. https://doi.org/10.1108/SEJ-09-2018-0060

[18] Knight, W. (2020) “Many Top AI Researchers Get Financial Backing From Big Tech” Wired. https://www.wired.com/story/top-ai-researchers-financial-backing-big-tech/.  

[19] Lin, HT., Balcan, MF., Hadsell, R., Razato, MA. (2020). “Reviewing is Underway!”. Medium. https://medium.com/@NeurIPSConf/reviewing-is-underway-a5532d4615ec. 

[20] Lowe, R. (2019). Introducing Retrospectives: ‘Real Talk’ for your Past Papers. The Gradient. https://thegradient.pub/introducing-retrospectives/

[21] Maybury, M. T. (1990). The mind matters: artificial intelligence and its societal implications. IEEE Technology and Society Magazine, 9(2), 7-15.

[22] National Science Foundation (2007), “Broader Impacts Review Criterion” https://www.nsf.gov/pubs/2007/nsf07046/nsf07046.jsp. 

[23] Neural Information Processing Systems Conference (NeurIPS) (2020). “Getting Started with NeurIPS 2020”. Medium. https://medium.com/@NeurIPSConf/getting-started-with-neurips-2020-e350f9b39c28. 

[24] Ovide, S. (2020). “A Case for Banning Facial Recognition” New York Times. https://www.nytimes.com/2020/06/09/technology/facial-recognition-software.html. 

[25] Partnership on AI. (2020) “Publication Norms for Responsible AI” https://www.partnershiponai.org/case-study/publication-norms/.

[26] Skrip, M. M. (2015). Crafting and evaluating Broader Impact activities: a theory‐based guide for scientists. Frontiers in Ecology and the Environment, 13(5), 273-279.

[27] TechTransformed (2019). “Consequence Scanning Agile Kit”. https://www.tech-transformed.com/product-development/. 

[28] Weiss, C. H. (1995). Nothing as practical as good theory: Exploring theory-based evaluation for comprehensive community initiatives for children and families. New approaches to evaluating community initiatives: Concepts, methods, and contexts, 1, 65-92.

\end{document}